\documentstyle[amsfonts,aps,prd]{revtex}

\begin{document}

\draft

\title{Collective Dynamics of Solitons and Inequivalent
Quantizations}

\author{Juan Pedro Garrahan}

\address{Departamento de F\'{\i}sica, Facultad de Ciencias Exactas y
Naturales, Universidad de Buenos Aires, Pabell\'on I, Ciudad
Universitaria, 1428 Buenos Aires, Argentina. \\
Email: garrahan@df.uba.ar}

\author{Mart\'{\i}n Kruczenski}

\address{Departamento de F\'{\i}sica, TANDAR, Comisi\'on Nacional de
Energ\'{\i}a At\'omica, Av. Libertador 8250, 1429 Buenos Aires,
Argentina. \\
Email: kruczens@tandar.cnea.edu.ar}

\date{October 9, 1997}

\maketitle

\begin{abstract}
The collective dynamics of solitons with a coset space $G/H$ as moduli
space is studied. It is shown that the collective band for a
vibrational state is given by the inequivalent coset space quantization
corresponding to the representation of $H$ carried by the vibration.
\end{abstract}


\section{Introduction}

Solitons arise as static solutions of finite energy to the equations of
motion of non linear field theories. A given solution in general
depends upon a set of parameters or moduli, and is a point in the
manifold of solutions of equal energy, or moduli space. In many cases
this manifold is simply a coset space $G/H$ where $G$ is the group of
symmetries of the action and $H \subset G$ is the symmetry of the
solitonic solution.

Around a soliton there are two kinds of quantum excitations, one
corresponding to collective motion in the moduli space, and the other
to vibrational excitations out of it. If the energy for the collective
excitations is much lower than that for the vibrational ones the low
energy spectrum can be approximately described by collective bands
associated with each vibrational state. 

Since the soliton is invariant under the subgroup $H$, vibrational
excitations fit into irreducible representations of $H$.  It is the
purpose of this letter to show that the collective band corresponding
to a vibrational state in a representation $\chi$ of $H$ realizes a
representation of $G$ induced by $\chi$. This representation is
reducible, so that when it is broken into irreducible representations
the whole collective band is obtained. This is equivalent to saying
that the collective band for a vibrational state is given by the
inequivalent quantization of $G/H$ corresponding to the representation
$\chi$ of $H$ carried by the vibration. In this way we show that
collective motion is a physical example of the inequivalent coset space
quantizations introduced by Mackey \cite{M69}, and more recently
studied by Landsman and Linden \cite{LL91} and MacMullan and Tsutsui
\cite{MT95}, among others.

\section{The model}

We consider a model \cite{GKB96} which formally corresponds to the
motion of a particle in a Riemannian manifold subject to a potential,
with action (summation over a repeated upper and lower index is 
implicit)
\begin{equation}
S = \int dt \; 
	\frac{1}{2} g_{st}(q) \dot{q}^s \dot{q}^t - V(q) .
	\label{eq:S} 
\end{equation}
The coordinates $q^s$ determine a configuration of the system. For
field theories, $q^s$ is a map from physical into
internal space, so the index $s$ stands for discrete internal indices
and continuous spatial coordinates, and sums over $s$ imply also integrals.

We assume that the action $S$ is invariant under an unbroken
finite-dimensional compact group $G$, of transformations of the fields
\begin{equation}
q \rightarrow R_g(q) , \;\;\; g \in G .
\end{equation}

Lets consider a static solution of the equations of motion $\bar{q}$
(i.e., a minimum of $V$), which is invariant only under a subgroup $H
\subset G$. In the case of a field theory $\bar{q}$ would be a soliton.
We will assume that the potential $V(q)$ is such that the moduli space
is given by the orbit of $\bar{q}$ under $G$, and is therefore a coset
space ${\cal M} = O_G(\bar{q}) = G/H$.

The fluctuations normal to ${\cal M}$ are massive or vibrational modes
and the tangent ones are zero or collective modes. We are interested in
the case in which the collective energy is much smaller than the
vibrational one.  Schematically this is given by $\hbar^2 / {\cal I}
\ll \hbar \omega \Rightarrow \hbar / {\cal I} \omega \ll 1$, where
${\cal I}$ is the inertia for the collective motion and $\omega$ the
frequency for the intrinsic excitations. This relation implies that it
is meaningful to expand in loops (in powers of $\hbar$).

One way to treat this system is to introduce collective coordinates by
performing a transformation of the fields $q$ with time dependent
parameters \cite{HK75,BK90,GKB96},
\begin{equation}
q \rightarrow R_{g(\alpha)}(q) ,
\end{equation}
where $\alpha^a(t)$ parameterize $G$. The action changes as
\begin{equation}
S \rightarrow \int dt \;  
	\frac{1}{2} g_{st}(q) D_0 q^s D_0 q^t - V(q) ,
	\label{eq:Sg}
\end{equation}
where the covariant derivatives are given by $D_0 q^s = \dot{q}^s +
\dot{\alpha}^a \zeta_a^b(\alpha) \delta_b q^s$, and $\zeta_a^b$ are
components of the Cartan-Maurer one-form $g^{-1} dg = d\alpha^a
\zeta_a^b T_b$, with $T_a$ being the group generators.  The transformed
action is invariant under gauge transformations of the fields $q$ and
the collective coordinates
\begin{equation}
\delta_{\varepsilon} q^s = \varepsilon^a(t) \delta_a q^s , \;\;\;
	\delta_{\varepsilon} \alpha^a = 
		- \varepsilon^b(t) \Theta_b^a(\alpha) , \;\;\; 
	( \Theta = \zeta^{-1} ) .
	\label{eq:gtrans}
\end{equation}

The action (\ref{eq:Sg}) can be seen as describing the problem from an
arbitrary moving frame of reference, its motion given by the collective
coordinates. While the gauge symmetry (\ref{eq:gtrans}) acts from the
right on the collective coordinates, the original symmetry $G$ acts
from the left.

In \cite{GKB96} we quantized the action (\ref{eq:Sg}) by means of the
Antifield formalism and calculated two-loops perturbative corrections
to the energies of the intrinsic states. In this letter, we are
interested in understanding the collective dynamics of an intrinsic
state in terms of quantization of free motion in the moduli space
${\cal M} = G/H$.

\section{Inequivalent Quantizations of $G/H$}

It was shown by Mackey \cite{M69} that when the configuration space of
a system is a coset space $G/H$ there are many different quantizations
not equivalent to each other by unitary transformations, which are
labeled by the irreducible unitary representations $\chi$ of the
subgroup $H$. The wave functions in a given inequivalent quantization
are vector valued, taking values in the representation space
$V_{\chi}$.\footnote{Wave functions are sections of a vector bundle
over configuration space \cite{LL91}.} They can be obtained from vector
valued functions in $G$ which satisfy ($\chi$-equivariant)
\begin{equation}
f_{(\chi\mu)}(g h) = 
	\sum_{\nu}
	\pi^{\chi}_{\nu \mu}(h) f_{(\chi\nu)}(g) ,
	\label{eq:mackey_condition}
\end{equation}
where $g \in G$, $h \in H$, and $\pi^{\chi}(h)$ are the matrices of the
representation $\chi$ of $H$.  Eq. (\ref{eq:mackey_condition}) implies
that the functions $f_{(\chi \mu)}(g)$ transform under the left action
of $G$ in the representation of $G$ induced by $\chi$. See
\cite{M69,BR86} for the definition and properties of induced
representations.

Landsman and Linden \cite{LL91} studied the dynamical consequences of
the inequivalent quantizations for the motion of a particle in $G/H$.
They discovered that in the non-trivial quantum sectors the particle
couples to a background gauge field $A_{\alpha}$, known as the
$H$-connection, which takes values in the representation of the
subalgebra $\pi^{\chi}({\frak h})$. The Hamiltonian is given by
\begin{equation}
{\cal H} = - \frac{1}{2} g^{\alpha \beta} 
	\left( \nabla_{\alpha} + A_{\alpha} \right) 
	\left( \partial_{\beta} + A_{\beta} \right) ,
	\label{eq:llham}
\end{equation}
where $\nabla_{\alpha}$ is the covariant derivative constructed out
of the metric $g_{\alpha \beta}$ on $G/H$.  Due to the $H$-connection
the Hamiltonian is matrix valued (in the trivial representation of $H$
it reduces to minus one half the Laplacian $- \frac{1}{2} \Delta_{G/H}
= - \frac{1}{2} g^{\alpha \beta} \nabla_{\alpha} \partial_{\beta}$).
An equivalent approach but with scalar wave functions and Hamiltonian
was developed by McMullan and Tsutsui \cite{MT95}, which allows for a
simpler definition of the corresponding path integral.

\section{Collective Dynamics}

We will now find out to which of the inequivalent quantizations
corresponds the collective band of a given intrinsic state. To see this
we consider the canonical quantization of the gauge system given by the
action (\ref{eq:Sg}).

The gauge transformations (\ref{eq:gtrans}) are generated by first
class constraints \cite{HT92}
\begin{equation}
\Phi_a = J_a - I_a , \label{eq:gens}
\end{equation}
where $J_a$ are the generators for the transformations $\delta_a q$ of
the intrinsic fields, and $I_a$ are the right generators for the
collective coordinates 
\begin{equation}
I_a = - i \Theta_a^b \frac{\partial}{\partial \alpha^b} .
\end{equation}
The generators $J_a$ can be divided into $J_i$ ($i =
1,\ldots,\mbox{dim}(H)$), which leave $\bar{q}$ fixed and so generate
the Lie algebra ${\frak h}$ of $H$, and $J_{\alpha}$ ($\alpha =
1,\ldots,\mbox{dim}(G/H)$), which generate the
complementary space ${\frak g-h}$. Since $H$ is a subgroup, the
operators $J_i$ close under commutation $[J_i,J_j] = i C_{ij}^k J_k$,
where $C_{ab}^c$ are the structure constants of $G$. Furthermore, as
$G$ is compact the generators $J_{\alpha}$ can be chosen in such a way that
\begin{equation}
{[}J_i,J_{\alpha}] = i C_{i\alpha}^{\beta} J_{\beta} ,
\label{eq:zerorep}
\end{equation} 
which shows that $J_{\alpha}$ transform in a representation (possibly 
reducible) of $H$.

We expand the fields $q$ in terms of fluctuations around $\bar{q}$
\begin{equation}
q^s = \bar{q}^s + 
	\sum_{\chi\mu} \psi_{(\chi \mu)}^s \hat{q}_{(\chi \mu)} +
	\sum_{\alpha} \psi_{\alpha}^s \hat{q}_{\alpha} ,
	\label{eq:qmodes}
\end{equation}
where $\psi_{(\chi \mu)}^s$ are the non-zero frequency normal modes of
the quadratic Hamiltonian, which, since $H$ maps the tangent space of
$\bar{q}$ into itself, fit into irreducible representations $\chi$ of
$H$. Different states in the irrep $\chi$ are labeled by $\mu = 1,
\ldots, \mbox{dim}(\chi)$. In the third term $\psi_{\alpha}^s =
\delta_{\alpha} {\bar{q}}^s$ are the zero modes, which also transform in
a representation of $H$ given by the structure constants
$C_{i\alpha}^{\beta}$ (see Eq. (\ref{eq:zerorep})). To leading order
the Hamiltonian reads
\begin{equation}
{\cal H} = \sum_{\chi \mu} 
	\omega_{\chi} \left( a_{(\chi \mu)}^{\dag} a_{(\chi \mu)} + 
	\frac{1}{2} \right) + 
	\frac{1}{2} {\cal I}^{\alpha \beta} p_{\alpha} p_{\beta} + 
	\cdots ,
\end{equation}
where ${\cal I}^{\alpha \beta}$ is the inverse of the inertia tensor
\begin{equation}
{\cal I}_{\alpha \beta} = 
	g_{st}(\bar{q}) \psi_{\alpha}^s \psi_{\beta}^t = 
	g_{st}(\bar{q}) \delta_{\alpha} \bar{q}^s 
	\delta_{\beta} \bar{q}^t ,
\end{equation}
which is invariant under the action of $H$ given by
(\ref{eq:zerorep}).

In terms of the fluctuations the intrinsic generators $J_a$ read
\begin{eqnarray}
J_{\alpha} &=& p_{\alpha} + 
	\sum_{nm} (D_{\alpha})_{n m} p_n \hat{q}_m + \cdots , 
	\label{eq:jgh} \\
J_i &=& \sum_{nm} (D_i)_{n m} p_n \hat{q}_m + \cdots .
	\label{eq:jh}
\end{eqnarray}
The first term in (\ref{eq:jgh}) shows that $J_{\alpha}$ generate the
zero modes $\delta_{\alpha} \bar{q}^s = \psi_{\alpha}^s$. The operators
$(p_n,\hat{q}_n)$ transform under $G$ in the representation given by
the constant matrices $D_{mn}$, where $n$ stands both for zero and
massive modes. The dots in the equations above stand for higher order
terms, which are present in the case of non linear transformations.

To eliminate the zero modes we choose $\hat{q}_{\alpha} = 0$ as a gauge
fixing for the constraints $\Phi_{\alpha}$ in (\ref{eq:gens}).
Substituting the commutators by Dirac brackets amounts to solving
$p_{\alpha}$ from the equations $\Phi_{\alpha} = 0$ \cite{HT92}
\begin{equation}
p_{\alpha} = I_{\alpha} - J_{\alpha}^{(2)} + \cdots ,
\end{equation}
where $J_{\alpha}^{(2)} = \sum_{nm} (D_{\alpha})_{nm} p_n \hat{q}_m$,
with $n$ and $m$ restricted to the finite modes $(\chi \mu)$. The
Hamiltonian reads
\begin{equation}
{\cal H} =\sum_{\chi\mu}  \omega_{\chi} 
	\left( a_{(\chi \mu)}^{\dag} a_{(\chi \mu)}
	+ \frac{1}{2} \right) + 
	\frac{1}{2} {\cal I}^{\alpha \beta} 
	( I_{\alpha} - J_{\alpha}^{(2)} ) 
	( I_{\beta} - J_{\beta}^{(2)} ) + \cdots .
	\label{eq:Hintcoll}
\end{equation}

A state in the collective band of a given intrinsic state carrying a
representation $\chi$ of $H$ (e.g., $a_{(\chi \mu)}^{\dag}|0\rangle$) has 
a wave function $\psi_{\chi}(q,\alpha)$. This wave function is restricted 
to satisfy the $H$-gauge invariance condition $\Phi_{i} \psi_{\chi} = 0$. 
This implies, together with (\ref{eq:gens}) and (\ref{eq:jh}), that
$\psi_{\chi}$ must be of the form
\begin{equation}
\psi_{\chi}(q,\alpha) = 
	\sum_{\mu} \varphi_{(\chi \mu)}(q) f_{(\chi \mu)}(\alpha), 
	\label{eq:state1}
\end{equation}
where $\varphi_{\chi \mu}(q) = \langle q | \chi \mu \rangle$ are the
intrinsic wave functions, and the collective functions $f_{(\chi \mu)}$
must satisfy
\begin{equation}
I_{i} f_{\chi \mu}(\alpha) = 
	i \sum_{\nu}
	(T_{i})_{\nu \mu}^{\chi} f_{(\chi \nu)}(\alpha) .
	\label{eq:fcoll}
\end{equation}
Here $(T_i)^{\chi}_{\mu \nu}$ are the generators of $H$ in the $\chi$
representation. This is the infinitesimal version of Eq.
(\ref{eq:mackey_condition}). A similar condition holds for the discrete
collective transformations $h \in H$, which together with
(\ref{eq:fcoll}) imply for all $h \in H$:
\begin{equation}
R_h f_{(\chi \mu)}(\alpha) = 
	\sum_{\nu}
	\pi^{\chi}_{\nu \mu}(h) f_{(\chi \nu)}(\alpha) .
	\label{eq:mackey2}
\end{equation}
Therefore, the collective functions $f_{(\chi \nu)}$ satisfy Mackey's
condition (\ref{eq:mackey_condition}), i.e., they transform under the
left action of $G$ in the representation of $G$ induced by the
representation $\chi$ of $H$ \cite{BR86}. In other words, the states in
the collective band are those of the inequivalent quantization of $G/H$
given by the representation $\chi$ of $H$ carried by the intrinsic
state.

By Peter-Weyl theorem $\psi_{\chi}(q,\alpha)$ in (\ref{eq:state1}) can
be written as \cite{BR86}
\begin{equation}
\psi_{\chi}(q,\alpha) = 
	\sum_{IMN \mu} \varphi_{(\chi\mu)} C^I_{MN (\chi \mu)} 
	D^I_{MN}(\alpha) ,
	\label{eq:fcoll2}
\end{equation}
where the sum is over all irreps $I$ of $G$ defined by the matrices
$D^I_{MN}(\alpha)$. 
Under the left action of $G$ each term of the sum transforms in the
corresponding representation $I$ of $G$, whereas under the right action
of $H$ transforms in the representation $I$ of $G$ considered as a
representation of $H$. This representation of $H$ is in general
reducible and can be broken in irreducible pieces $\chi^I_k$. Condition
(\ref{eq:mackey2}) implies that only those $k$ such that
$\chi^I_k=\chi$ should be kept, i.e., $f_{(\chi\mu)}(\alpha)$ must
transform under the representation $\chi$ of $H$. Hence, we rewrite
(\ref{eq:fcoll2}) as
\begin{equation}
\psi_{\chi}(q,\alpha) = 
	\sum_{IM} C^I_{M \chi} \psi_{IM \chi}(q,\alpha) = 
	\sum_{IM, \; k/\chi_k=\chi, \; \mu} C^I_{M \chi k}  
	\varphi_{(\chi\mu)}(q) D^I_{M (\chi_k, \mu)}(\alpha) ,
	\label{eq:wavefunction}
\end{equation}
where the index $N$ was replaced by its decomposition into $H$ irreps
$(\chi_k,\mu)$. This means that each representation $I$ of $G$ appears
in the collective band as many times as the representation $\chi$ of
$H$ is contained in the decomposition of $I$ into $H$ irreps (we call
this number $d^I_{\chi}$).  As the collective Hamiltonian, given by the
second term in Eq. (\ref{eq:Hintcoll})
\begin{equation}
{{\cal H}}_{\rm coll} =\frac{1}{2} {\cal I}^{\alpha \beta} 
	( I_{\alpha} - J_{\alpha}^{(2)} ) 
	( I_{\beta} - J_{\beta}^{(2)} ) ,
	\label{eq:Hcoll}
\end{equation}
does not mix different representations $I$, it can be diagonalized in
subspaces of dimension $d^I_{\chi}$. It is easy to check that it also
commutes with the simultaneous action of $H$ on the intrinsic states
($J_i$) and on the collective coordinates from the right ($I_i$). This
ensures that it preserves the physical condition $\Phi_i=0$ which the
states $\psi_{\chi}$ satisfy.

In perturbation theory we must restrict the intrinsic part of ${\cal
H}_{\rm coll}$ to the given intrinsic subspace of dimension
$\mbox{dim}(\chi)$. Indicating this restriction with an overbar, ${\cal
H}_{\rm coll}$ becomes:
\begin{equation}
{{\cal H}}_{\rm coll} =
	\frac{1}{2} {\cal I}^{\alpha \beta} I_{\alpha} I_{\beta} - 
	{\cal I}^{\alpha \beta}  I_{\alpha}  
	\overline{J_{\beta}^{(2)}} +
	\frac{1}{2} {\cal I}^{\alpha \beta} 
	\overline{ J_{\alpha}^{(2)} J_{\beta}^{(2)}} .
	\label{eq:Hcoll'}
\end{equation}
The second term is a ``Coriolis'' term, and the third one is constant.
The diagonalization of ${\cal H}_{\rm coll}$ in each subspace of
dimension $d^I_{\chi} > 1$ must be performed case by case. However,
some more information can be obtained using the $H$-invariance. The
restricted matrices $\overline{J_{\alpha}^{(2)}}$ no longer satisfy the
$G$ algebra, but they still satisfy $[J_i,\overline{J_{\alpha}^{(2)}}]
=C_{i\alpha}^{\beta} \overline{J_{\beta}^{(2)}}$,
which means that they transform under $H$ in the same representation as
$J_{\alpha}$. By the Wigner-Eckart theorem in the group $H$, they are
determined by Clebsch-Gordan coefficients up to as many independent
constants as irreps of $H$ are contained in this representation. The
number of irreps also gives the number of independent inertia moments
in ${\cal I}_{\alpha \beta}$. 

If $G/H$ is a symmetric space a considerable simplification arises because
$\overline{J_{\alpha}^{(2)}} = 0$. This can be shown using
the involutive automorphism which defines the symmetric space, under
which $H$ is invariant and the generators $J_{\alpha}$ change sign
\cite{BR86}. Another simplification is that there is only one moment of
inertia, so the collective Hamiltonian becomes 
\begin{equation}
{{\cal H}}_{\rm coll} = 
	\frac{1}{2 {\cal I}} I_{\alpha}^2 = 
	\frac{1}{2 {\cal I}} \left( I_a^2 - I_i^2 \right) ,
\end{equation}
and the $d^I_{\chi}$ states $\psi_{IM \chi}$ in (\ref{eq:wavefunction})
are degenerate, since $I_a^2$ and $I_i^2$ are Casimirs of $G$ and $H$,
respectively.

If we choose $\alpha^{i} = 0$ to fix the $H$-gauge invariance, the
collective functions become $f_{(\chi \mu)}(x)$, with $x \in G/H_0$, where
$H_0$ is the identity component of $H$. The
derivatives respect to $\alpha^i$ are replaced by solving $\Phi_i = 0$,
which yield
\begin{eqnarray}
I_i &=& J_i = - i \sum_{\chi \mu \nu}
	(T_{i})^{\chi}_{\mu \nu} p_{(\chi \mu)} \hat{q}_{(\chi \nu)} + 
	\cdots , \\
I_{\alpha} &=& 
	- i \Theta_{\alpha}^{\beta}(x) 
	\left( \partial_{\beta} - 
	\zeta_{\beta}^i(x)
 	\sum_{\chi \mu \nu} (T_{i})^{\chi}_{\mu \nu} p_{(\chi \mu)} 
              \hat{q}_{(\chi \nu)} \right) + \cdots .
\end{eqnarray}
Written in the gauge $\alpha^i=0$ the collective Hamiltonian becomes
(up to a constant term)
\begin{equation}
{\cal H}_{\rm coll} = - \frac{1}{2} g^{\alpha \beta} 
	\left( \nabla_{\alpha} + A_{\alpha} \right) 
	\left( \partial_{\beta} + A_{\beta} \right) - \frac{1}{2}
	g^{\alpha \beta} \left[
	\left( \nabla_{\alpha} + A_{\alpha} \right) \tilde{A}_{\beta} +
	\tilde{A}_{\alpha} \left( \partial_{\beta} + A_{\beta} \right) 
	\right] .
\label{eq:hllcoll}
\end{equation}
The first term corresponds to Landsman-Linden Hamiltonian
(\ref{eq:llham}), with the metric on $G/H_0$ given by $g_{\alpha \beta} =
{\cal I}_{\gamma \delta} \zeta_{\alpha}^{\gamma} \zeta_{\beta}^{\delta}$,
and the $H_0$-connection $(A_{\alpha})_{\mu \nu} = - i \zeta_{\alpha}^{i}
(T_{i})^{\chi}_{\mu \nu}$. The second term gives the coupling
to an extra connection $(\tilde{A}_{\alpha})_{\mu \nu} = - i
\zeta_{\alpha}^{\beta} (\overline{J_{\beta}^{(2)}})_{\mu \nu}$ which
comes from the Coriolis terms. In general, for real representations,
this is an $SO(\mbox{dim}\,\chi)$ connection, and is similar to the
induced connections studied in \cite{M95}.

The $H_0$-connection ensures that Hamiltonian (\ref{eq:hllcoll})
applied to functions independent of $\alpha_i$ gives the same result as
Hamiltonian (\ref{eq:Hcoll'}) acting on functions over $G$ which satisfy
(\ref{eq:fcoll}).  When $H_0\neq H$ the functions $f_{(\chi \mu)}(x)$ are
still restricted by condition (\ref{eq:mackey2}) for the discrete elements
of $H$. This restriction can be lifted including a pure gauge connection
which associates a discrete element of $H$ with each non-trivial path
in $\Pi_1(G/H)$ (holonomy factors).

\section{Conclusions}

We have discussed the collective bands of intrinsic states found when
quantizing around a soliton solution with moduli space isomorphic to
$G/H$. The result is that the collective band of an intrinsic vibrational
state realizes an inequivalent coset space quantization given by the
representation of $H$ under which the intrinsic state transforms. The
collective Hamiltonian is that of Landsman and Linden \cite{LL91},
which describes free motion on $G/H$ coupled to a background $H$-gauge
field. Besides, there may be other background gauge fields coming from
the Coriolis terms.  The extra degrees of freedom associated with the
non-trivial quantizations are given by the intrinsic coordinates. In this
way, we have given a physical example of the inequivalent quantizations
studied in \cite{M69,LL91,MT95}.

This work may be of interest for obtaining the spins and isospins of
the ground state of multiskyrmions \cite{BTC90,BS97} and their excited
states (which have been found for topological charges $B=2$ and $B=4$
by Barnes et al. \cite{BBT97a,BBT97b}). The symmetry group of the
Skyrme model is $G = O(3)_{\rm spin} \times O(3)_{\rm isospin}$,
and the elements of $H \subset G$ are of the form $h = (\tilde{h},
D(\tilde{h}))$, with $\tilde{h} \in \tilde{H} \subset O(3)$, and $D :
\tilde{h} \rightarrow D(\tilde{h})$ a three dimensional orthogonal
representation of $\tilde{H}$ \cite{LM94}. For $B > 2$ the subgroup $H$
is discrete, and the problem of determining the collective bands is
analogous to obtaining the rotational spectra of polyatomic molecules
\cite{LL65}.

\acknowledgements

We are grateful to N.S. Manton, D.R. Bes, B.J. Schroers, and N.N.
Scoccola for enlightening discussions. JPG has been supported by Consejo
Nacional de Investigaciones Cient\'{\i}ficas y T\'ecnicas, Argentina
(CONICET). Partial support from Fundaci\'on Antorchas is acknowledged.

\end{document}